\documentclass[aps,pre,reprint,nofootinbib]{revtex4-2}

\usepackage{amsmath,amssymb,bm}

\begin{document}

\title{Jarzynski equality for counterwork under reversed memory-filtered driving}

\author{Pierre Nazé}
\email{pnaze@ufpa.br}
\affiliation{
Universidade Federal do Par\'a, ICEN, Faculdade de F\'isica,
Av. Augusto Corr\^ea, 1, Guam\'a, 66075-110, Bel\'em, Par\'a, Brazil
}

\date{\today}

\begin{abstract}
We introduce a counterwork functional generated by a sign-inverting
memory-filtered effective protocol. Given an imposed protocol $\lambda(t)$,
the effective protocol $\Lambda(t)$ is obtained by applying an active
protocol-memory kernel to $\dot{\lambda}(t)$, rather than by invoking a
passive bath response. The counterwork is then the ordinary Hamiltonian work
associated with $H(\Gamma,\Lambda(t))$, so that Jarzynski's equality applies
directly to it. When $\Lambda(t)$ reverses the endpoints of $\lambda(t)$, the
corresponding free-energy difference satisfies $\Delta F_C=-\Delta F_W$, and
the exponential average of the counterwork is the reciprocal of that of the
original work. We derive the kernel normalization realizing this reversed
displacement and show, by Jensen's inequality, that the product of the
exponentials of the average work and counterwork is bounded by unity,
implying $\langle C\rangle+\langle W\rangle\geq0$. Thus negative average
counterwork is possible only when compensated by the average work of the
original operation.  We further discuss the counteroperation under incomplete thermodynamic information, showing that the robust strategy is to enforce endpoint reversal while minimizing dissipated counterwork.
\end{abstract}

\maketitle

\textit{Quando acordou, Alfredo estava nu e sem memória. Traído, sem herança ou herdeiros, e com suas ideias e honra roubadas, perdera tudo contra todos. Lembrara-se de Jó e de suas infelicidades. Decidiu, então, escrever sua nova história.
\vspace{-0.2cm}
\begin{flushright}
--- As Infelicidades de Alfredo
\end{flushright}}
\vspace{-0.3cm}

\section{Introduction}

Fluctuation relations provide exact constraints on the statistics of work
performed on systems driven far from equilibrium~\cite{Jarzynski1997,Crooks1999,Campisi2011,Seifert2012}. Among them, Jarzynski's
equality relates the exponential average of the work performed during a
finite-time transformation to the equilibrium free-energy difference between
the initial and final values of the externally controlled parameter. For a
Hamiltonian system driven by a protocol $\lambda(t)$, the work is defined by
the parametric variation of the Hamiltonian, and the equality holds provided
the system is initially prepared in the corresponding canonical equilibrium
state.

In this work we consider a complementary construction based on an effective
protocol $\Lambda(t)$ obtained from the imposed protocol $\lambda(t)$ through
a memory-filtered velocity. The memory kernel considered here is not a
passive response function of the thermal bath. Rather, it is interpreted as
part of the external driving apparatus, specifying how the imposed protocol is
mapped into an effective one. This distinction allows for sign-changing
kernels, which may represent active feedback or inversion mechanisms.

The central object is a counterwork functional $C$, defined as the ordinary
Hamiltonian work generated by the effective protocol $\Lambda(t)$. Since $C$
is work associated with the Hamiltonian family $H(\Gamma,\Lambda)$,
Jarzynski's equality applies to it in the standard way, provided the initial
state is canonical with respect to $H(\Gamma,\Lambda(0))$. When the effective
protocol reverses the endpoints of the original protocol, namely
$\Lambda(0)=\lambda(\tau)$ and $\Lambda(\tau)=\lambda(0)$, the free-energy
difference associated with the counterwork satisfies
$\Delta F_C=-\Delta F_W$. Consequently, the exponential average of the
counterwork is the reciprocal of the exponential average of the original
work.

This relation does not imply that the counterwork is equal to minus the
original work along individual trajectories. Instead, it relates two
fluctuation relations associated with two different canonical preparations:
the original work process starts at $\lambda(0)$, whereas the counterwork
process starts at $\Lambda(0)=\lambda(\tau)$. If the two processes are
interpreted as consecutive stages of a single operation, an equilibration step
at fixed parameter must be inserted between them. This relaxation contributes
heat but no work, and prepares the canonical state required for the Jarzynski
equality associated with the counterwork.

A further point concerns the asymmetry between the original operation and the
counteroperation. The original protocol fixes the final value of the control
parameter, while the counteroperation starts only after this endpoint has
already been imposed. Thus the counteroperator receives the system at the
final value selected by the original driving and constructs a reversed
effective protocol from that point. In general, the counteroperator need not
know in advance the sign of the free-energy difference of the original
process. Since $\Delta F_C=-\Delta F_W$, the counteroperation may either be
work-extracting or work-consuming, depending on the free-energy landscape. The
robust thermodynamic strategy is therefore not to prescribe the sign of the
counterwork, but to enforce endpoint reversal while minimizing the dissipated
counterwork.

The purpose of the paper is therefore twofold. First, we identify the kernel
condition under which a memory-filtered driving map realizes a reverse
thermodynamic transformation. Second, we clarify the thermodynamic status of
the corresponding counteroperation under incomplete information about the
free-energy difference. The construction remains consistent with the second
law because the sign-inverting kernel changes the effective driving protocol;
it does not alter the thermodynamic bound applied to the process generated by
$H(\Gamma,\Lambda(t))$. A complete energetic description of the full cycle,
including heat exchanged during equilibration stages and internal energy
changes, is left for future work.

\section{Original work}

Consider a classical system coupled to a heat bath with inverse temperature $\beta$. The total Hamiltonian is
\begin{equation}
H=H(\Gamma,\lambda),
\end{equation}
where $\Gamma$ denotes a phase-space point and $\lambda$ is an externally controlled
parameter. Let the original protocol be
\begin{equation}
\lambda(t)=\lambda_0+g(t)\delta\lambda,
\qquad
g(0)=0,
\qquad
g(\tau)=1.
\end{equation}
Therefore,
\begin{equation}
\lambda(0)=\lambda_0,
\qquad
\lambda(\tau)=\lambda_0+\delta\lambda.
\end{equation}

The usual work performed along a realization is
\begin{equation}
W
=
\int_0^\tau
\partial_\lambda H(\Gamma_t,\lambda(t))\dot\lambda(t)\,dt.
\end{equation}
Throughout the paper, work is defined with the Jarzynski convention:
positive work corresponds to work performed on the system by the external
driving apparatus. The equilibrium partition function is
\begin{equation}
Z(\lambda)
=
\int d\Gamma\, e^{-\beta H(\Gamma,\lambda)}.
\end{equation}
The corresponding free energy is
\begin{equation}
F(\lambda)
=
-\frac{1}{\beta}\ln Z(\lambda).
\end{equation}
Thus the free-energy difference associated with the original work is
\begin{equation}
\Delta F_W
=
F(\lambda_0+\delta\lambda)-F(\lambda_0),
\end{equation}
or, equivalently,
\begin{equation}
\Delta F_W
=
-\frac{1}{\beta}
\ln
\frac{
Z(\lambda_0+\delta\lambda)
}{
Z(\lambda_0)
}.
\end{equation}
Assuming that the system is initially prepared in the canonical state associated
with $H(\Gamma,\lambda_0)$, Jarzynski's equality gives
\begin{equation}
\left\langle e^{-\beta W}\right\rangle
=
e^{-\beta\Delta F_W}.
\end{equation}

\section{Counterwork generated by a reversed effective protocol}

We now introduce an effective protocol $\Lambda(t)$ generated by a memory-filtered
velocity,
\begin{equation}
\dot\Lambda(t)
=
\int_0^\tau K(t,t')\dot\lambda(t')\,dt'.
\end{equation}
Here, $K(t,t')$ is not assumed to be a passive response function of the bath.
It is a protocol-memory kernel specifying how the externally imposed
driving $\lambda(t)$ is mapped into the effective driving $\Lambda(t)$.
Therefore, sign-changing kernels are allowed when the driving apparatus
contains active feedback or inversion. The effective protocol is
\begin{equation}
\Lambda(t)
=
\Lambda(0)
+
\int_0^t ds
\int_0^\tau
K(s,t')\dot\lambda(t')\,dt'.
\end{equation}

The counterwork is defined as the Hamiltonian work generated by the effective
protocol:
\begin{equation}
C
=
\int_0^\tau
\partial_\Lambda H(\Gamma_t,\Lambda(t))\dot\Lambda(t)\,dt.
\end{equation}
The same sign convention is used for the counterwork $C$. Thus, negative values of $C$ correspond to work extracted from the system by the effective reversed driving. Substituting the memory-filtered velocity, one obtains
\begin{equation}
C
=
\int_0^\tau dt\,
\partial_\Lambda H(\Gamma_t,\Lambda(t))
\int_0^\tau dt'\,
K(t,t')\dot\lambda(t').
\end{equation}

This definition makes $C$ an ordinary Hamiltonian work, but associated with the
effective protocol $\Lambda(t)$ rather than with the imposed protocol $\lambda(t)$.
The relevant Hamiltonian family is therefore $H(\Gamma,\Lambda)$.

\section{Free-energy difference associated with counterwork}

The partition function associated with $H(\Gamma,\Lambda)$ is
\begin{equation}
Z(\Lambda)
=
\int d\Gamma\, e^{-\beta H(\Gamma,\Lambda)}.
\end{equation}
The free energy is
\begin{equation}
F(\Lambda)
=
-\frac{1}{\beta}\ln Z(\Lambda).
\end{equation}
The free-energy difference associated with the counterwork is
\begin{equation}
\Delta F_C
=
F(\Lambda(\tau))-F(\Lambda(0)).
\end{equation}
Equivalently,
\begin{equation}
\Delta F_C
=
-\frac{1}{\beta}
\ln
\frac{
Z(\Lambda(\tau))
}{
Z(\Lambda(0))
}.
\end{equation}

Assume that the system is initially prepared in the canonical state associated
with $H(\Gamma,\Lambda(0))$,
\begin{equation}
\rho_0^C(\Gamma)
=
\frac{
e^{-\beta H(\Gamma,\Lambda(0))}
}{
Z(\Lambda(0))
}.
\end{equation}
Since $C$ is the work generated by $H(\Gamma,\Lambda(t))$, Jarzynski's equality gives
\begin{equation}
\left\langle e^{-\beta C}\right\rangle
=
e^{-\beta\Delta F_C}.
\end{equation}

\section{Condition for opposite free-energy differences}

We now impose that the effective protocol reverses the endpoints of the original
protocol:
\begin{equation}
\Lambda(0)=\lambda(\tau)=\lambda_0+\delta\lambda,
\qquad
\Lambda(\tau)=\lambda(0)=\lambda_0.
\end{equation}
Using the reversed endpoints,
\begin{equation}
\Delta F_C
=
F(\lambda_0)-F(\lambda_0+\delta\lambda).
\end{equation}
Therefore,
\begin{equation}
\Delta F_C
=
-\left[
F(\lambda_0+\delta\lambda)-F(\lambda_0)
\right].
\end{equation}
Since
\begin{equation}
\Delta F_W
=
F(\lambda_0+\delta\lambda)-F(\lambda_0),
\end{equation}
one obtains
\begin{equation}
\Delta F_C=-\Delta F_W.
\end{equation}

Thus the Jarzynski equality for the counterwork becomes
\begin{equation}
\left\langle e^{-\beta C}\right\rangle
=
e^{\beta\Delta F_W}.
\end{equation}
Using
\begin{equation}
\left\langle e^{-\beta W}\right\rangle
=
e^{-\beta\Delta F_W},
\end{equation}
we also have
\begin{equation}
\left\langle e^{-\beta C}\right\rangle
=
\frac{1}{
\left\langle e^{-\beta W}\right\rangle
}.
\end{equation}
Equivalently,
\begin{equation}
\left\langle e^{-\beta C}\right\rangle
\left\langle e^{-\beta W}\right\rangle
=
1.
\end{equation}
Remark that the averages in the two fluctuation relations refer to different initial canonical preparations: the original work process starts at $\lambda_0$, whereas the counterwork process starts at $\Lambda(0)=\lambda_0+\delta\lambda$. If the two processes are interpreted as consecutive stages of a single
operation, an equilibration step must be inserted after the forward work.
After the protocol $\lambda(t)$ ends at $\lambda_0+\delta\lambda$, the
system is generally out of equilibrium. It must therefore be allowed to
relax at fixed parameter value $\lambda_0+\delta\lambda$ before the
counterwork process starts. Since the parameter is kept fixed during this
relaxation, this step contributes no work, although heat may be exchanged
with the bath. It prepares the canonical initial state required for the
Jarzynski equality associated with $C$.

\section{Exponential bound for the average work and counterwork}

The fluctuation relations obtained for the original work and for the
counterwork involve exponential averages,
\begin{equation}
\left\langle e^{-\beta W}\right\rangle
=
e^{-\beta\Delta F_W},
\end{equation}
and
\begin{equation}
\left\langle e^{-\beta C}\right\rangle
=
e^{-\beta\Delta F_C}.
\end{equation}
When the effective protocol reverses the endpoints of the original protocol,
one has $\Delta F_C=-\Delta F_W$, and therefore
\begin{equation}
\left\langle e^{-\beta C}\right\rangle
\left\langle e^{-\beta W}\right\rangle
=
1 .
\end{equation}
This identity concerns the product of exponential averages. It should be
distinguished from the product of exponentials of the average work and
counterwork.

By Jensen's inequality,
\begin{equation}
e^{-\beta\langle W\rangle}
\leq
\left\langle e^{-\beta W}\right\rangle ,
\end{equation}
and
\begin{equation}
e^{-\beta\langle C\rangle}
\leq
\left\langle e^{-\beta C}\right\rangle .
\end{equation}
Multiplying these two inequalities and using the product relation for the
fluctuation averages gives
\begin{equation}
e^{-\beta\langle C\rangle}
e^{-\beta\langle W\rangle}
\leq
\left\langle e^{-\beta C}\right\rangle
\left\langle e^{-\beta W}\right\rangle
=
1 .
\end{equation}
Equivalently,
\begin{equation}
e^{-\beta(\langle C\rangle+\langle W\rangle)}
\leq
1 .
\end{equation}
Since $\beta>0$, this implies
\begin{equation}
\langle C\rangle+\langle W\rangle\geq 0 .
\end{equation}

This inequality gives a useful interpretation of negative average
counterwork. If $\langle C\rangle<0$, then
$-\beta\langle C\rangle>0$, and the factor
$e^{-\beta\langle C\rangle}$ is larger than unity. The product bound can then remain satisfied only if the factor
$e^{-\beta\langle W\rangle}$ is sufficiently small. Equivalently, the
average original work must satisfy
$\langle W\rangle\geq -\langle C\rangle$. Thus, if the negative average
counterwork is large in units of $k_B T$, the compensating average work must
also be large in the same units.

\section{Kernel condition for reversed displacement}

The effective displacement is
\begin{equation}
\delta\Lambda
=
\Lambda(\tau)-\Lambda(0).
\end{equation}
From the definition of $\Lambda(t)$,
\begin{equation}
\delta\Lambda
=
\int_0^\tau dt
\int_0^\tau dt'\,
K(t,t')\dot\lambda(t').
\end{equation}
Changing the order of integration,
\begin{equation}
\delta\Lambda
=
\int_0^\tau dt'\,
\dot\lambda(t')
\left[
\int_0^\tau K(t,t')\,dt
\right].
\end{equation}
The original displacement is
\begin{equation}
\delta\lambda
=
\lambda(\tau)-\lambda(0)
=
\int_0^\tau \dot\lambda(t')\,dt'.
\end{equation}
To reverse the endpoints, one needs
\begin{equation}
\delta\Lambda=-\delta\lambda.
\end{equation}
Therefore,
\begin{equation}
\int_0^\tau dt'\,
\dot\lambda(t')
\left[
\int_0^\tau K(t,t')\,dt
\right]
=
-\int_0^\tau \dot\lambda(t')\,dt'.
\end{equation}
For arbitrary protocol velocity $\dot\lambda(t')$, this is guaranteed by
\begin{equation}
\int_0^\tau K(t,t')\,dt=-1,
\qquad
0\leq t'\leq \tau.
\end{equation}
This is the sign-reversing analogue of the endpoint-preserving normalization.
The normalization is imposed over the output time $t$, because it fixes the total
displacement of the effective protocol.

\section{Example}

Consider the constant sign-inverting kernel
\begin{equation}
K(t,t')=-\frac{1}{\tau}.
\end{equation}
It satisfies
\begin{equation}
\int_0^\tau K(t,t')\,dt=-1.
\end{equation}

Take the original protocol to be linear,
\begin{equation}
\lambda(t)=\lambda_0+\frac{t}{\tau}\delta\lambda.
\end{equation}
Then
\begin{equation}
\dot\lambda(t)=\frac{\delta\lambda}{\tau}.
\end{equation}
The effective velocity is
\begin{equation}
\dot\Lambda(t)
=
\int_0^\tau
\left(
-\frac{1}{\tau}
\right)
\frac{\delta\lambda}{\tau}
\,dt'
=
-\frac{\delta\lambda}{\tau}.
\end{equation}
Choose
\begin{equation}
\Lambda(0)=\lambda_0+\delta\lambda.
\end{equation}
Then
\begin{equation}
\Lambda(t)
=
\lambda_0+\delta\lambda-\frac{t}{\tau}\delta\lambda.
\end{equation}
Thus
\begin{equation}
\Lambda(\tau)=\lambda_0.
\end{equation}
The effective protocol therefore runs from the final value of the original protocol
back to its initial value:
\begin{equation}
\Lambda(0)=\lambda(\tau),
\qquad
\Lambda(\tau)=\lambda(0).
\end{equation}
Consequently,
\begin{equation}
\Delta F_C
=
F(\lambda_0)-F(\lambda_0+\delta\lambda)
=
-\Delta F_W.
\end{equation}
The corresponding fluctuation relation is
\begin{equation}
\left\langle e^{-\beta C}\right\rangle
=
e^{\beta\Delta F_W}
=
\frac{1}{
\left\langle e^{-\beta W}\right\rangle
}.
\end{equation}
Although the example uses a linear protocol, the sign-reversing
normalization condition is protocol-independent: for arbitrary
$\dot\lambda(t')$, it enforces $\delta\Lambda=-\delta\lambda$.

\section{Second-law consistency}

The sign-reversing kernel does not violate the second law. It changes the effective
driving protocol. The second law must be applied to the process actually generated
by $H(\Gamma,\Lambda(t))$.

From Jarzynski's equality for $C$,
\begin{equation}
\left\langle e^{-\beta C}\right\rangle
=
e^{-\beta\Delta F_C}.
\end{equation}
By Jensen's inequality,
\begin{equation}
e^{-\beta\langle C\rangle}
\leq
\left\langle e^{-\beta C}\right\rangle.
\end{equation}
Therefore,
\begin{equation}
e^{-\beta\langle C\rangle}
\leq
e^{-\beta\Delta F_C}.
\end{equation}
Since $\beta>0$, this implies
\begin{equation}
\langle C\rangle\geq \Delta F_C.
\end{equation}
If $\Delta F_C=-\Delta F_W$, then
\begin{equation}
\langle C\rangle\geq -\Delta F_W.
\end{equation}
The dissipated counterwork is
\begin{equation}
C_{\rm diss}
=
\langle C\rangle-\Delta F_C,
\end{equation}
and it satisfies
\begin{equation}
C_{\rm diss}\geq 0.
\end{equation}

Thus the sign-reversing kernel is compatible with the second law, provided it is
interpreted as an active inversion of the driving protocol rather than as a passive
positive memory kernel. In particular, defining the dissipated work as
\begin{equation}
    W_{\rm diss} = \langle W\rangle-\Delta F_W,
\end{equation}
one obtains
\begin{equation}
    \langle W\rangle+\langle C\rangle=W_{\rm diss}+C_{\rm diss}\ge 0,
\end{equation}
attaining $\langle C\rangle=-\langle W\rangle$ only when both the forward
and reversed effective processes are quasistatic. In other words, the average total work of the forward and counterwork stages cannot be negative.

There is another limiting situation in which both the work and the counterwork vanish, and hence their average sum also vanishes for finite-time processes. If the Hamiltonian does not depend on the externally controlled parameter,
\begin{equation}
H(\Gamma,\lambda)=H(\Gamma),
\end{equation}
then the parameter variation does not couple to the system. Hence,
\begin{equation}
\partial_\lambda H(\Gamma,\lambda)=0,
\end{equation}
and the ordinary work vanishes,
\begin{equation}
W=
\int_0^\tau
\partial_\lambda H(\Gamma_t,\lambda(t))\dot\lambda(t)\,dt
=0.
\end{equation}
Since the effective protocol $\Lambda(t)$ is only a memory-filtered
reparametrization of $\lambda(t)$, the Hamiltonian also satisfies
\begin{equation}
H(\Gamma,\Lambda)=H(\Gamma),
\qquad
\partial_\Lambda H(\Gamma,\Lambda)=0.
\end{equation}
Therefore the counterwork also vanishes,
\begin{equation}
C=
\int_0^\tau
\partial_\Lambda H(\Gamma_t,\Lambda(t))\dot\Lambda(t)\,dt
=0.
\end{equation}
Consequently, the partition function and the free energy are independent of
both $\lambda$ and $\Lambda$, implying
\begin{equation}
\Delta F_W=0,
\qquad
\Delta F_C=0.
\end{equation}
Thus the fluctuation relations become trivial,
\begin{equation}
\left\langle e^{-\beta W}\right\rangle
=
\left\langle e^{-\beta C}\right\rangle
=
1.
\end{equation}
In this case, the kernel may reverse the protocol, but it cannot generate
work or counterwork unless the Hamiltonian actually depends on the controlled
parameter.

\section{Counteroperation under incomplete thermodynamic information}

The preceding construction assumes that the effective protocol $\Lambda(t)$
reverses the endpoints of the original protocol $\lambda(t)$. This means that
the original operator fixes the final value of the control parameter through
the forward driving,
\begin{equation}
\lambda(0)=\lambda_0,
\qquad
\lambda(\tau)=\lambda_0+\delta\lambda .
\end{equation}
The counteroperation starts only after this final value has been imposed.
Therefore, the counteroperator does not choose the state from which the
counterwork process starts. Rather, the counteroperator receives the system
at the endpoint of the original operation and defines an effective protocol
satisfying
\begin{equation}
\Lambda(0)=\lambda(\tau),
\qquad
\Lambda(\tau)=\lambda(0).
\end{equation}
In this sense, the counteroperation is conditioned by the outcome of the
original operation.

A relevant point is that the counteroperator need not know in advance the
sign of the free-energy difference associated with the original process.
Since
\begin{equation}
\Delta F_C=-\Delta F_W ,
\end{equation}
the sign of the counterwork free-energy difference depends on the sign of
$\Delta F_W$. If $\Delta F_W>0$, then $\Delta F_C<0$, and the lower bound for
the average counterwork is negative,
\begin{equation}
\langle C\rangle\geq -\Delta F_W .
\end{equation}
In this case, the average counterwork may be negative, corresponding to work
extracted from the system by the reversed effective driving. Conversely, if
$\Delta F_W<0$, then $\Delta F_C>0$, and
\begin{equation}
\langle C\rangle\geq \Delta F_C>0 .
\end{equation}
Thus the counteroperator must invest positive work on average. Therefore, in
the absence of information about the sign of $\Delta F_W$, the counteroperator
cannot generally decide whether the counteroperation will be work-extracting
or work-consuming.

The robust thermodynamic strategy is therefore not to optimize the sign of the
counterwork, but to minimize the dissipated counterwork,
\begin{equation}
C_{\rm diss}
=
\langle C\rangle-\Delta F_C
\geq 0 .
\end{equation}
This is achieved, in principle, by implementing the reversed effective
protocol as close as possible to a quasistatic transformation. The
counteroperator should therefore choose a sign-inverting memory map that
enforces the endpoint reversal,
\begin{equation}
\int_0^\tau K(t,t')\,dt=-1,
\end{equation}
while avoiding unnecessary finite-time dissipation. The role of the
counteroperation is then not to force a prescribed energetic sign, but to
realize the reverse thermodynamic transformation with the least possible
excess work.

This interpretation also clarifies the physical meaning of the active
sign-inverting kernel. The kernel does not determine whether the
counteroperation is energetically favorable or costly; that information is
contained in the free-energy landscape through $\Delta F_C$. The kernel only
specifies the external map that converts the imposed protocol into a reversed
effective protocol. Hence, the counteroperator acts as a reverser of the
endpoint displacement, not as an agent that can bypass the thermodynamic
bound. The second law remains encoded in
\begin{equation}
\langle C\rangle\geq \Delta F_C .
\end{equation}
The optimal counteroperation is therefore a controlled reversal with minimal
dissipation, rather than an attempt to guarantee work extraction.

\section{Conclusion}

We have introduced a counterwork functional associated with a
memory-filtered effective protocol. Given an imposed protocol $\lambda(t)$,
the effective protocol $\Lambda(t)$ is generated by applying a
protocol-memory kernel to the velocity $\dot{\lambda}(t)$. The kernel is
interpreted as part of the external driving apparatus, rather than as a
passive thermal response function. This interpretation allows
sign-changing kernels, including sign-inverting kernels that implement an
active reversal of the imposed displacement.

The counterwork $C$ was defined as the ordinary Hamiltonian work generated by
the effective protocol $\Lambda(t)$ through the Hamiltonian family
$H(\Gamma,\Lambda)$. Therefore, Jarzynski's equality applies to $C$ in the
standard way, provided the initial state is canonical with respect to
$H(\Gamma,\Lambda(0))$. When the effective protocol reverses the endpoints of
the original protocol, the associated free-energy difference satisfies
\begin{equation}
\Delta F_C=-\Delta F_W .
\end{equation}
Consequently, the exponential average of the counterwork is the reciprocal of
the exponential average of the original work,
\begin{equation}
\left\langle e^{-\beta C}\right\rangle
=
\frac{1}{\left\langle e^{-\beta W}\right\rangle}.
\end{equation}
This relation concerns exponential averages and does not imply that
$C=-W$ trajectory by trajectory. It also does not imply an equality between
the ordinary averages $\langle C\rangle$ and $\langle W\rangle$.

Applying Jensen's inequality to the two fluctuation relations gives the bound
\begin{equation}
e^{-\beta\langle C\rangle}
e^{-\beta\langle W\rangle}
\leq
1 .
\end{equation}
Thus negative average counterwork is thermodynamically allowed, but only if
it is compensated by the average work performed during the original
operation. In particular, if $\langle C\rangle<0$, then
$\langle W\rangle\geq-\langle C\rangle$. The compensation is linear at the
level of average energies, although it appears exponentially in the product
of exponentials of the averages.

We also identified the condition under which the memory-filtered protocol
realizes a reversed displacement. For arbitrary protocol velocity, endpoint
reversal is guaranteed by the normalization condition
\begin{equation}
\int_0^\tau K(t,t')\,dt=-1,
\qquad
0\leq t'\leq\tau .
\end{equation}
The constant kernel $K(t,t')=-1/\tau$ gives the simplest example, producing a
linear effective protocol that runs from the final value of the original
protocol back to its initial value.

The construction is consistent with the second law because the latter must
be applied to the actual effective process generated by
$H(\Gamma,\Lambda(t))$. Introducing the dissipated work and dissipated
counterwork, one obtains
\begin{equation}
\langle W\rangle+\langle C\rangle
=
W_{\rm diss}+C_{\rm diss}
\geq0 .
\end{equation}
Equality is reached only when both the original and reversed effective
processes are thermodynamically reversible. In the limiting case in which
the Hamiltonian is independent of the control parameter, both $W$ and $C$
vanish, the free-energy differences are zero, and the fluctuation relations
become trivial.

Finally, the counteroperation was analyzed under incomplete thermodynamic
information. The counteroperator receives the endpoint imposed by the
original driving and need not know in advance the sign of $\Delta F_W$.
Consequently, the reversed operation may be work-extracting or
work-consuming, depending on the free-energy landscape. The robust strategy
is therefore not to prescribe the sign of the counterwork, but to enforce
endpoint reversal while minimizing dissipated counterwork. In this sense, the
sign-inverting memory kernel acts as a controlled reversal mechanism: it does
not bypass the thermodynamic bound, but specifies an active map from the
imposed protocol to an effective protocol with reversed displacement.

A complete energetic description of the full operation should also include
the heat exchanged during the equilibration stages, the corresponding
internal energy changes, and the entropy production of the full cycle. These
aspects are left for future work.

\end{document}